\documentstyle[bo99,epsfig]{article}

\title{X-ray Beaming in the High Magnetic Field Pulsar GX 1+4}
\author{D.K. Galloway $^{1,2}$ and K. Wu $^{2}$ }
\affil{ $^1$ School of Mathematics and Physics,
       University of Tasmania,
       GPO Box 252-21, Hobart,
       Tasmania 7001, Australia \\
  $^2$ Special Research Centre for Theoretical Astrophysics,
       University of Sydney,
       Camperdown, NSW 2006, Australia }

\begin{document}

\maketitle

\begin{abstract}
Pulse profiles from X-ray pulsars often exhibit strong energy dependence and
both periodic and aperiodic variations with time. The great variety of
profiles observed in various sources, and even from individual sources,
makes it difficult to separate the numerous factors influencing the
phase-dependence of the X-ray emission. These factors include the system
geometry and particularly the photon energy and angle dependence of emission
about the neutron star poles.

Comptonisation may play an important role in determining beam patterns and
hence pulse profiles in X-ray pulsars. A Monte Carlo simulation is used to
investigate the beaming due to Comptonisation in a simple accretion column
geometry.  We apply the model to the extremely variable pulse profiles of
the high-magnetic field pulsar GX 1+4.
\end{abstract}

\section{Introduction}

To date essentially all the approaches which have been used to model the
emission region in X-ray pulsars have limitations. Past efforts have
typically adopted a geometry suitable for a particular accretion rate
(${\dot M}$) regime and then predict the emission properties by a range of
techniques.  Radiative transfer calculations (e.g.  Burnard, Arons \& Klein
1991) are necessarily restricted to symmetric, homogeneous emission regions,
where in reality the accretion column may be hollow and even incomplete (an
`accretion curtain'). The geometric fitting approach, where a beam pattern
is assumed and then the geometry is varied (e.g. Leahy 1991) cannot
reproduce sharper features observed in several sources.  Neither method can
generate asymmetric pulse profiles without resorting to an off-center
magnetic axis, for which there is no other observational evidence.  Recent
observations of the X-ray pulsar GX~1+4 suggest a rather different scenario.

The X-ray continuum spectrum of GX~1+4 is rather flat (with photon index
$\approx 1.0$) up to a cutoff around 10-20~keV, above which the decay is
steeper; it is one of the hardest known amongst the X-ray pulsars.  Analysis
of recent Rossi X-ray Timing Explorer ({\it RXTE}) data shows that the
spectrum is generally consistent with those predicted by unsaturated
Comptonisation models (e.g.  Galloway et al. 1999).  Pulse profiles are
extremely variable and typically asymmetric, often with a sharp dip forming
the primary minimum (Greenhill, Galloway \& Storey 1998).  During a low flux
episode in July 1996, the pulse profiles were found to shift in asymmetry
from `leading-edge bright' (with the maximum closely following the sharp
primary minimum) to `trailing-edge bright'. The entire observation which
captured the change spanned only 34 hours, and occurred just 10 days before
a short-lived transition from rather constant spin-down to spin-up and back
again (Giles et al. 1999). We propose a model which seeks to explain the
sharp primary minima seen in this and other sources (A 0535+262, Cemeljic \&
Bulik 1998; and RX J0812.4-3114, Reig \& Roche 1999) and ultimately the
change in the pulse profiles.

\section{Model Description and Preliminary Results}

A Monte-Carlo code is used to generate the spectra and pulse profiles
emitted by two semi-infinite homogeneous cyclindrical accretion columns of
radius $R_C$, diametrically located on the surface of a `canonical' neutron
star of radius $R_*=10$~km (Figure 1a).

The algorithms of Pozdnyakov, Sobol' \& Syunyaev (1983) are used to draw the
photon energy and direction, electron energies, and to calculate the fully
relativistic (non-magnetic) cross-section for Compton scattering.  Outside
the accretion column, the redshift and bending of photon trajectories by the
neutron star's gravity is calculated by assuming a Schwarzschild metric.  We
simulate a single column for both poles of the star, and generate pulse
profiles for a range of geometries simultaneously.

 \begin{figure}
 \centerline{\psfig{file=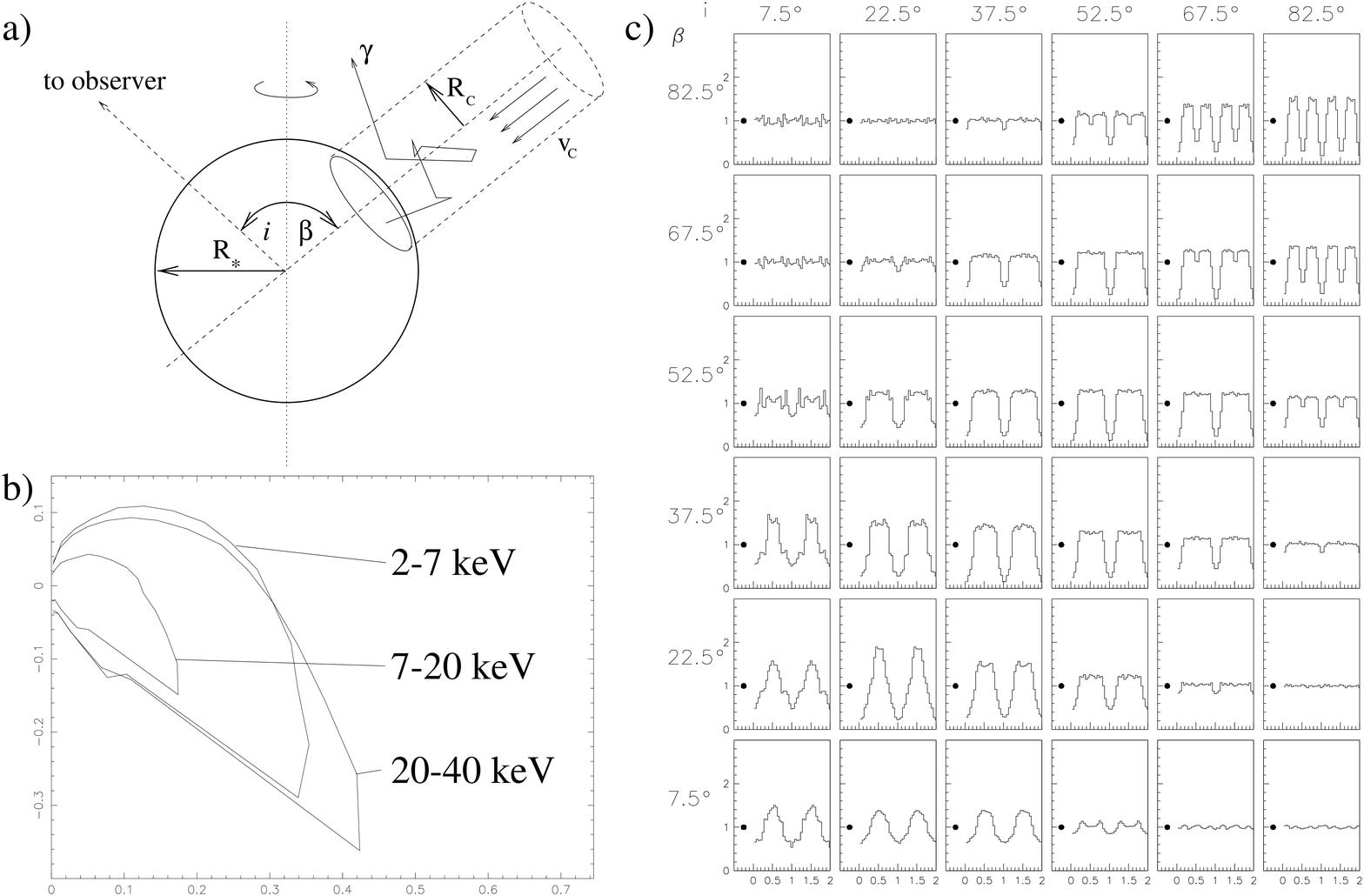, width=13cm}}
\vspace{-0.7cm}
 \caption[]{a) Model geometry (only one column shown for clarity).  The
magnetic axis of the star is aligned relative to the rotational axis by an
angle $\beta$, with $i$ the inclination angle of the system with respect to
the observer.  Photons are emitted isotropically from each (circular) polar
cap, and are then Compton scattered by the column plasma before escaping
towards the observer.  The plasma is flowing towards the pole with speed
$v_C$.  The temperature of the polar cap $T_0$, temperature $T_e$ and
optical depth $\tau$ of the column plasma, and column diameter $R_C$ are all
free parameters in the model. Note that $\tau=\sigma_{\rm T}R_C N_e$ where
$N_e$ is the electron density in the column; that is, $\tau$ corresponds to
photon paths from the centre of the column radially outwards. Clearly the
optical depth experienced by individual photons will depend on the
trajectory, and in particular with the angle relative to the column axis.
\\
b) X-ray beam pattern within several energy bands.  The model parameters are
$kT_0=1$~keV, $kT_e=8$~keV, $\tau=3$, $R_C=1$~km, and $v_C=0.5c$; these
conditions are such as to approximately correspond to those expected for a
low-luminosity X-ray pulsar.  The origin corresponds to the base of the
accretion column, which is aligned along the (positive) $y$-axis. The
normalisation is arbitrary.\\
c) Predicted pulse profiles using the same model parameters over a range of
geometries. Each profile shown corresponds to a particular choice of $i$ and
$\beta$, which vary between $7.5^\circ$ and $82.5^\circ$ along the $x$- and
$y$-axes respectively.  Profiles are normalised to the mean and plotted over
two cycles; a typical error bar is shown at the left of each panel.
}
 \end{figure}

The beam pattern and pulse profiles over a range of geometries are shown in
Figure 1 b) and c).  We note that when ($|i-\beta|\la 45^\circ$) the
emission exhibits a strong modulation at the stars rotation period, with
the primary minimum corresponding to the closest passage of the line of
sight with one of the magnetic polar axes.  As $i$ and $\beta$ increase the
primary minimum becomes progressively narrower. When
$i\approx\beta\ga50^\circ$, a secondary minimum (from the passage of the
second axis through the line of sight) is observed.  The emission is beamed
at an angle $>90^\circ$ with respect to the column axis; this corresponds to
a `fan' type beam.  Emission at smaller angles is supressed as a consequence
of the decreased escape probability for photons propagating along the column
axis.  That emission is beamed at $>90^\circ$ is a consequence of the
gravitational light bending; this is also affected by the size of the column
$R_C$. 

Our simulations indicate that the mean spectra also depend strongly on the
density of the column and the viewing geometry.

\section{Discussion and application to X-ray pulsars}

Since we assume a constant infall velocity $v_C$ and neglect effects due to
radiation pressure on the infalling electrons, the results described are
only applicable to systems with low ${\dot M}$.  Previous low-${\dot M}$
models predict a `pencil' rather than `fan' beam, with emission reaching a
maximum at small angles relative to the accretion column. This is probably
because of the assumed `slab' or `mound' shaped emission region.
Interactions between photons and the inflowing material in the accretion
column, which is neglected by these models, is crucial for the formation of
sharp primary minima in the pulse profiles as observed in GX~1+4 and several
X-ray pulsars. The persistence of the sharp feature in GX~1+4 as X-ray flux
drops almost to zero points to the continued importance of this effect, even
at extremely low ${\dot M}$ (Giles et al. 1999).

A significant approximation is the use of the nonmagnetic Compton scattering
cross-section.  For GX~1+4 - with an estimated magnetic field strength of
$2-3 \times 10^{13}$~G (Cui et al. 1997) - deviations from the nonmagnetic
cross section will be significant within typical observational bands for
X-ray astronomy. However we suspect that magnetic effects may only play a
minor role in shaping the pulse profile, principally narrowing the primary
minimum and possibly giving rise to the local maxima (`shoulders')
immediately prior to and following the minimum (Giles et al. 1999).

Finally we note that the model-predicted pulse profiles are in general quite
symmetric. A possible cause of asymmetry in the observed profiles is a
variation in density across the accretion column, which could potentially
develop in the region where the disc plasma becomes entrained onto the
magnetic field lines and persist to the neutron star surface.  This effect
further suggests a mechanism for the rapid changes in profile asymmetry
observed in GX~1+4 (Giles et al. 1999); that is, the sense of asymmetry in
the column changes and consequently so does the pulse profile. The detailed
structure of the entrainment region is rather poorly understood, and we feel
it is not possible to rule out such a phenomenon.

We have described a model with homogeneous, axisymmetric, cyclindrical
emission regions. To explain other qualitative features of observed pulse
profiles, our model needs to be modified to take into account effects due to
inhomogeneities and more complicated geometry of the emission regions.


\end{document}